\documentstyle[twoside,12pt,epsf]{article}
\normalsize
\newcommand{\bdi}{\begin{displaymath}}
\newcommand{\edi}{\end{displaymath}}
\newcommand{\bfi}{\begin{figure}}
\newcommand{\efi}{\end{figure}}

\newcommand{\beq}{\begin{equation}}
\newcommand{\eeq}{\end{equation}}
\newcommand{\beqa}{\begin{eqnarray}}
\newcommand{\eeqa}{\end{eqnarray}}

\newcommand{\ra}{\rightarrow}

\newcommand{\wt}{\widetilde}

\def\longbar#1{\setbox1=\hbox{$#1$}
\setbox2=\vbox{\hrule width 0.8\wd1}
\raise0.5\ht1\hbox{${\lower\dp1\box2}\atop\box1$}}  

\begin{document}

\begin{titlepage}

\begin{flushright}
KA--TP--24--1999 \\
\today
\end{flushright}

\vspace{1cm}
\begin{center}
{\Large \bf Hopf instantons and the Liouville equation in target space}\\[1cm]
C. Adam* \\
Institut f\"ur Theoretische Physik, Universit\"at Karlsruhe \\

\medskip

\medskip

B. Muratori**,\, C. Nash*** \\
Department of Mathematical Physics, National University of Ireland, Maynooth
\vfill
{\bf Abstract} \\
\end{center}
We generalise recent results on Hopf instantons in a Chern--Simons \& Fermion
theory in a fixed background magnetic field. We find that these instanton 
solutions have to obey the Liouville equation in target space. As a
consequence, these solutions are given by a class of Hopf maps that 
consist of the composition of the standard Hopf map with an arbitrary
rational map.

\vfill

$^*)${\footnotesize  
email address: adam@maths.tcd.ie, adam@pap.univie.ac.at} 

$^{**})${\footnotesize
email address: bmurator@fermi1.thphys.may.ie} 

$^{***})${\footnotesize
email address: cnash@stokes2.thphys.may.ie} 
\end{titlepage}

\section{Introduction}

In this letter, we shall generalise some recent results \cite{hoin}
on fully three-dimensional solutions of an Abelian Chern--Simons and
Fermion system in the presence of a fixed background magnetic field.
These solutions are related to Hopf maps and are, therefore, labelled 
by the Hopf index.
Hopf maps are just maps $S^3 \ra S^2$. These maps fall into
different homotopy classes that are labelled by the integers (the Hopf
index, see below for details). 
Field theories  where static solutions with
nontrivial Hopf index (Hopf solitons) occur have already received considerable
interest recently (see e.g., \cite{Ran1}--\cite{AFZ}).

Chern--Simons theories 
have been widely studied ever since their introduction \cite{DJT}.
Specifically,
when an Abelian Chern--Simons term in three dimensions is coupled to matter,
the magnetic field is forced to be proportional to the electric current
due to the equations of motion \cite{JLW}--\cite{Dun1}. 
Further, in these models there
exist soliton-like, static (i.e., two-dimensional) solutions that are
related to some topological invariants (e.g., maps $S^2 \ra S^2$) 
\cite{JLW}--\cite{Dun1}. 
Usually, these solitons behave like vortices, and, because of their 
topological nature, they exhibit magnetic flux quantization.
Therefore these solutions are physically relevant in situations
where the phenomenon of magnetic flux quantization occurs and where
matter is confined to a plane, the most prominent example being the
quantum Hall effect \cite{ZHK,Jain}.

At this point the question arises whether there exist fully
three-dimensional solutions for such Chern--Simons \& matter systems,
and whether these solutions may be characterized by some topological
invariants, as well.   

In a recent paper \cite{hoin} we have demonstrated that,
if the presence of a 
fixed, prescribed 
background magnetic field is assumed, then there indeed exist solutions to the
Chern--Simons \& Fermion system defined below. 
These solutions are related to the specific Hopf maps
\beq
S^3 \stackrel{\chi}{\ra} S^2 \stackrel{R_n}{\ra}S^2
\eeq
where $\chi $ is the standard Hopf map with Hopf index 1 and 
\beq
R_n : z\ra z^n \, , \quad z \in {\rm\bf C}\, ,\quad n\in {\rm\bf Z}
\eeq
is a special class of rational maps ${\rm\bf C}\ra {\rm\bf C}$ that express
maps $S^2 \ra S^2$ in stereographic coordinates (see below for details).
The Hopf maps (1) have Hopf index $N=n^2$, and $n$ is the winding number 
of the corresponding map $S^2 \ra S^2$, (2).

In this letter we shall show that the special class (2) of rational maps
in (1) may actually be generalised to arbitrary rational maps $R(z)$. 
These general rational maps will emerge as solutions of the Liouville
equation on the target $S^2$ in (1) that the Hopf instantons will be
proven to obey.

The paper is organised as follows. In Section 2 we briefly review some
features of maps $S^2 \ra S^2$ and of Hopf maps $S^3 \ra S^2$. In
Section 3, we define our Chern--Simons and Fermion system. We show that its
solutions in the presence of a fixed prescribed background magnetic field
are solutions to the Liouville equation in target space. Consequently,
these solutions are given by general rational maps in target space.
We discuss our results in the final section.

\section{Maps $S^2 \ra S^2$ and Hopf maps $S^3 \ra S^2$}

Maps $S^2 \ra S^2$ are characterised by their winding number $w$. One way 
of describing them is by interpreting both $S^2$ as Riemann spheres
and by introducing stereographic coordinates $z\in {\rm\bf C}$ on both
of them. A specific class of such maps $S^2 \ra S^2$ may then be described
by rational maps
\beq
R: \, z\ra R(z) =\frac{P(z)}{Q(z)}
\eeq
where $P(z)$ and $Q(z)$ are polynomials, and $z$ and $R(z)$ are interpreted as 
stereographic coordinates on the domain and target $S^2$, respectively. The
winding number $w$ of this map is given by the degree of the map,
\beq
w={\rm deg}(R)={\rm max}(p,q)
\eeq
where $p$ and $q$ are the degrees of the polynomials $P(z)$ and $Q(z)$.
Another possibility of computing the same winding number involves the
pullback under $R(z)$ of the standard area two-form $\Omega$ on $S^2$
(in stereographic coordinates),
\beq
\Omega =\frac{2}{i}\frac{d\bar z dz }{(1+z\bar z)^2} \, , \quad
\int \Omega =4\pi .
\eeq
The pullback is ($'$ means derivative w.r.t. the argument)
\beq
R^* \Omega =\frac{2}{i}\frac{|R' (z)|^2}{(1+R\bar R)^2}d\bar z dz
=: \frac{2}{i}\rho_R (z,\bar z)d\bar z dz
\eeq
and obeys
\beq
\int R^* \Omega =4\pi w
\eeq
where $w$ is again the winding number (4). For later convenience we want to 
point out that the function $\rho_R (z,\bar z)$ defined in (6) obeys the
Liouville equation for all rational maps $R(z)$ (in fact for all 
holomorphic functions $f(z)$; however, we will be forced to restrict to
rational maps later on by regularity requirements),
\beq
\partial_z \partial_{\bar z} \ln \rho_R (z,\bar z)= -2\rho_R (z,\bar z)
\eeq
where
\beq
\partial_z \partial_{\bar z}=\frac{1}{4}(\partial_x^2 + \partial_y^2)
=\frac{1}{4}\Delta
\eeq
($\Delta$ is the two-dimensional Laplacian) and we expressed $z$ by
its real and imaginary part,
\beq
z=x+iy \, ,\quad \partial_z =\frac{1}{2}(\partial_x - i\partial_y ).
\eeq
If we separate the function $\rho_z (z,\bar z)=(1+z\bar z)^{-2}$ that
corresponds to the identity map $z\ra z$ from $\rho_R (z,\bar z)$,
\beq
\rho_R =:\rho_z \wt\rho_R
\eeq
then $\wt\rho_R$ obeys the equation
\beq
\partial_z \partial_{\bar z}\ln \wt\rho_R =
-2\frac{\wt\rho_R -1}{(1+z\bar z)^2}
\eeq
which we shall need later on.

Hopf maps are maps $S^3 \ra S^2$. They may be expressed, e.g., by maps $\chi :
{\rm\bf R}^3 \ra {\rm\bf C}$ provided that the complex function $\chi$
obeys $\lim_{|\vec x|\ra \infty} \chi (\vec x)=\chi_0 ={\rm const}$,
where $\vec x =(x_1 ,x_2 ,x_3)^{\rm T}$. The pre-images in ${\rm\bf R}^3$
of points of the target $S^2$ (i.e., the pre-images of points $\chi =
{\rm const}$) are closed curves in ${\rm\bf R}^3$ (circles in the 
related domain $S^3$). Any two different circles are linked $N$ times,
where $N$ is the Hopf index of the given Hopf map $\chi$.
Further, a magnetic field $\vec B$ 
(the Hopf curvature) is related to the
Hopf map $\chi$ via
\beq
\vec B = \frac{2}{i}\frac{(\vec\partial\bar\chi)\times(\vec\partial\chi)}{
(1+\bar\chi \chi)^2} =4\frac{S(\vec\partial S)\times\vec
\partial\sigma}{(1+S^2)^2}
\eeq
where $\chi =Se^{i\sigma}$ is expressed in terms of its modulus $S$
and phase $\sigma$ at the r.h.s. of (13).

Mathematically, the curvature $F=\frac{1}{2}F_{ij}dx_i dx_j$, $F_{ij}=
\epsilon_{ijk}B_k$, is the pullback under the Hopf map, $F=\chi^* \Omega$,
of the standard area two-form $\Omega$ , (5), on the target $S^2$. 
Geometrically, $\vec B$ is tangent to the closed curves $\chi ={\rm
  const}$ (see e.g. \cite{Ran1,FN1,BS1,JaPi3}; 
the authors of \cite{JaPi3}
describe Hopf curvatures slightly
differently, by the Abelian projection of an $SU(2)$ pure gauge
connection, which has some technical advantages).
The Hopf index $N$ of $\chi$ may be computed from $\vec B$ via
\beq
N=\frac{1}{16\pi^2}\int d^3 x \vec A \vec B
\eeq
where $\vec B=\vec\partial\times \vec A$.

The simplest (standard) Hopf map $\chi$ with Hopf index $N=1$ is
\beq
\chi =\frac{2(x_1 +ix_2)}{2x_3 -i(1-r^2)}
\eeq
with modulus and phase
\beq
S^2 = \frac{4(r^2 -x_3^2)}{4x_3^2 +(1-r^2)^2}\, ,\quad
\sigma ={\rm atan}\, \frac{x_2}{x_1} + {\rm atan}\, \frac{1-r^2}{2x_3} 
\eeq
(a Hopf map has to be single valued, but may well be singular, as $\chi 
=\infty$ is just the south pole of the target $S^2$). 
This $\chi$, (15), leads to the Hopf curvature 
\beq
\vec B = \frac{16}{(1+r^2)^2} \vec N
\eeq
and we have introduced the unit vector
\beq
\vec N = \frac{1}{1+r^2}
\left( \begin{array}{c} 2x_1 x_3 -2x_2  \\ 2x_2 x_3 +2x_1 \\
1-x_1^2 -x^2_2 +x_3^2 \end{array} \right) 
\eeq
($\vec N^2 =1$) for later convenience.

\section{Construction of the Hopf instantons}

We start with the action ($i,j,k =1 \ldots 3$)
\beq
S=\int d^3 x \Bigl( \Psi^\dagger (-i\partial_j -\bar A_j)\sigma_j \Psi
+ \frac{1}{2}\vec A \vec B \Bigr)
\eeq
where $\Psi$ is a two-component spinor (Fermion), $\sigma_j$ are the
Pauli matrices and $\vec A$ is an Abelian gauge potential. Further,
\beq
S_{\rm CS}=
\frac{1}{2}\int d^3 x\vec A\vec B =\frac{1}{4}\int d^3 x 
\epsilon_{ijk}A_iF_{jk}
\eeq
is the Chern--Simons (CS) action, where the Chern--Simons coupling 
constant is chosen equal to one; and
\beq
\bar A_i = A_i + A_i^{\rm B}
\eeq
where the background gauge field $A_i^{\rm B}$ and its magnetic field
$B_i^{\rm B}=\epsilon_{ijk}\partial_j A^{\rm B}_k$ are
\beq
\vec A^{\rm B} 
=-\frac{1}{1+r^2} \vec N \, , \quad 
\vec B^{\rm B} =- \frac{4}{(1+r^2)^2} \vec N
\eeq
and  the unit vector $\vec N$ is given in (18).

Observe that the (fixed, non-dynamical) background field is coupled to the
Fermion, but it is absent in the CS term. The equations of motion resulting
from the action (19) are
\beq
(-i\partial_j -\bar A_j )\sigma_j \Psi =0,
\eeq
(the Dirac equation) and
\beq
\vec\Sigma := \Psi^\dagger \vec\sigma \Psi = \vec B .
\eeq
Observe that for any pair $(\Psi ,\bar A_j)$ that solves the Dirac equation 
(23) the spin density $\vec\Sigma$ related to $\Psi$ has to obey
\beq
\vec\partial \vec\Sigma =0,
\eeq
therefore, equations (23) and (24) are consistent.

The simplest solution to this system is (see \cite{LoYa,Zero})
\beq
\Psi =\frac{4}{(1+r^2)^\frac{3}{2}} ({\bf 1} +i\vec x \vec\sigma )
\left( \begin{array}{c} 1 \\ 0 \end{array} \right) 
\eeq
\beq
\vec A 
=\frac{4}{1+r^2} \vec N  
\eeq
\beq
\vec\Sigma =\vec B = \frac{16}{(1+r^2)^2} \vec N
\eeq
($\vec N$ is given in (18)).
Here the dynamical gauge field is proportional to the background field,
therefore one could find a solution without background field by choosing
either a different normalization of the fermion (26) or by choosing a
Chern--Simons coupling constant in (20), (24) different from 1.
However, this will not be true for the solutions below, for which the
background field (22) is crucial.

Further, the magnetic field (and spin density) of this simplest solution, (28),
is precisely equal to the Hopf curvature (17) of the simplest standard Hopf 
map, (15). Here the question arises whether there are more solutions to 
(23), (24)
that are characterised by Hopf maps, and we already know from \cite{hoin}
that this is indeed the case. Here we shall generalise these results,
therefore
 we should provide some more Hopf maps that will give rise to more
solutions to (23), (24). We will produce these Hopf maps by composing the
standard Hopf map with some maps $S^2 \ra S^2$, i.e.,
\beq
\chi_R :S^3 \stackrel{\chi^{(1)}}{\ra} S^2 \stackrel{R}{\ra} S^2
\eeq
where $R(z)$ is a general rational map (3).
Such Hopf maps have Hopf index $N = w^2$, where $w$ is the winding number
(i.e., the degree (4) of the rational map).

Next we need some facts about the Dirac equation (23). Suppose a spinor
$\Psi $ is given that solves (23) for some $\bar A_i$, then this
gauge field $\bar A_i$ may actually be 
expressed in terms of the zero mode $\Psi$ as \cite{LoYa}
\bdi
\bar A_i =\frac{1}{|\vec \Sigma |}(\frac{1}{2}\epsilon_{ijk}\partial_j
\Sigma_k +{\rm Im}\, \Psi^\dagger \partial_i \Psi )   
\edi
\beq 
= \frac{1}{2}\epsilon_{ijk}(\partial_j \ln |\vec \Sigma |){\cal N}_k
+\frac{1}{2}\epsilon_{ijk}\partial_j {\cal N}_k +
{\rm Im}\, \widehat\Psi^\dagger
\partial_i \widehat\Psi
\eeq
where we have expressed $\bar A_i$ in terms of the general
unit vector and unit spinor
\beq
\vec {\cal N}=\frac{\vec \Sigma}{|\vec \Sigma |}\, ,\quad \widehat\Psi =
\frac{\Psi}{|\Psi^\dagger \Psi |^{1/2}} .
\eeq
Now we are able to construct the solutions to (23), (24) as follows. We define
the spinor
\beq
\Psi^{(M)} = e^{i\Lambda}e^{M/2}\Psi
\eeq
where $\Psi$ at the r.h.s. of (32) is just the zero mode of the simplest
solution, (26). Further,  $M$ is a real function of the
simplest standard Hopf map $\chi$, (15), and its complex conjugate
$\bar\chi$. $\Lambda$ is a pure gauge factor that has to be chosen
accordingly (see below). 

For the corresponding spin density $\Sigma^{(M)}$ it still holds that
\beq
\Sigma^{(M)}_{i,i}=e^M ((M_{,\chi}\chi_{,i}+M_{,\bar\chi}\bar\chi_{,i} )
\Sigma_{,i} +\Sigma_{i,i})=0
\eeq
where $\Sigma_i$ is the spin density (and magnetic field) in (28), or
equivalently the Hopf curvature (17) of the standard Hopf map (15).
Therefore, $\Psi^{(M)}$ is still a zero mode. The corresponding gauge field
$\bar A^{(M)}_i$ that solves the Dirac equation
together with $\Psi^{(M)}$ may be computed with the help of (30) to be
\beq
\bar A^{(M)}_i =\bar A_i +\frac{1}{2}\epsilon_{ijk}(\partial_j M)
N_k +\Lambda_{,i}
\eeq
where $\bar A_i$ is the gauge field (27) of the simplest solution
plus the background gauge
field (22), and $\vec N$ is the specific unit vector (18).
The corresponding magnetic field $\bar B^{(M)}_i = \epsilon_{ijk}
\partial_j \bar A^{(M)}_k$ is
\bdi
\bar B^{(M)}_l =\bar B_l +\frac{1}{2}[M_{,\chi} 
(\chi_{,lk}N_k +\chi_{,l}N_{k,k}
-\chi_{,kk}N_l - \chi_{,k}N_{l,k})+
\edi
\bdi
M_{,\bar\chi} 
(\bar\chi_{,lk}N_k +\bar\chi_{,l}N_{k,k}
-\bar\chi_{,kk}N_l - \bar\chi_{,k}N_{l,k})
\edi
\beq
-(M_{,\chi\chi}\chi_{,k}\chi_{,k} +M_{,\bar\chi \bar\chi}\bar\chi_{,k}
\bar\chi_{,k} +2M_{,\chi \bar\chi}\chi_{,k}\bar\chi_{,k}) N_l]
\eeq
where $\bar B_l$ is the magnetic field (28) plus the background 
magnetic field (22). 
After some tedious algebra we find that only the coefficient of
$M_{,\chi \bar\chi}$ is nonzero, i.e.,
\beq
\chi_{,lk}N_k +\chi_{,l}N_{k,k}
-\chi_{,kk}N_l - \chi_{,k}N_{l,k} =0
\eeq
\beq
\chi_{,k}\chi_{,k} =0
\eeq
\beq
\chi_{,k}\bar\chi_{,k} = 8\frac{(1+\chi\bar\chi)^2}{(1+r^2)^2}
\eeq
and, therefore
\beq
\bar B^{(M)}_l = \bar B_l -8\frac{(1+\chi\bar\chi)^2}{(1+r^2)^2}
M_{,\chi\bar\chi}N_l .
\eeq
Now we should insert this into the Chern--Simons equation (24) 
after subtracting the background magnetic field (22),
\beq
B^{(M)}_l =\bar B^{(M)}_l - B^{\rm B}_l =\Sigma^{(M)}_l .
\eeq
We arrive at
\beq
\frac{16}{(1+r^2)^2}N_l - \frac{8(1+\chi\bar\chi)^2}{(1+r^2)^2}
M_{,\chi\bar\chi}N_l =\frac{16e^M}{(1+r^2)^2}N_l
\eeq
or
\beq
M_{,\chi\bar\chi}=-2\frac{e^M -1}{(1+\chi\bar\chi)^2}
\eeq
which is just version (12) of the Liouville equation (for $\wt\rho =
(1+z\bar z)^2 \rho$). However, equation (42) holds in target space
(i.e., for ``coordinates'' $\chi ,\bar\chi$).
Solutions to (42) are therefore
\beq
M=\ln \Bigl( (1+\chi\bar\chi)^2 \frac{|R' (\chi)|^2}{(1+R(\chi )
\bar R(\chi))^2} \Bigr)
\eeq
for arbitrary rational functions $R(\chi)$.

Here we still have to explain why the restriction to rational maps
$R(\chi)$ is necessary, because in principle any holomorphic function
$f(\chi)$ in (43) solves eq. (42).
This question is related to the pure gauge factor $\Lambda$ in (32) and (34),
which we have not yet determined, because it did not show up in eq. (42),
which is gauge invariant.

The point is that the gauge field $\bar A^{(M)}_l$, as defined in (34), is 
singular at the zeros of $\exp (M(\chi ,\bar\chi))$. 
In other words, $\bar A^{(M)}_l$
is singular along closed curves that are pre-images $\chi (\vec x)
=z_i$ of the zeros of $\exp (M)$. The $M$-dependent part of 
$\bar A^{(M)}_l$ in (34) may be rewritten as
\beq
\frac{1}{2}\epsilon_{ljk}M_{,j}N_k = \frac{i}{2}(M_{,\chi}\chi_{,l}
- M_{,\bar\chi}\bar\chi_{,l})
\eeq
as may be checked easily. Now assume that $M$ is given by (43) for some
$R(\chi) =P(\chi)/Q(\chi)$, then the zeros of $\exp (M)$ are the zeros
of
\beq
|R'(\chi)|^2 = |PQ' - P' Q|^2
\eeq
and, generically, there are $p+q-1$ zeros (if we assume $p>q$, which we may,
see below, then there are precisely $p+q-1$ zeros). Each individual
zero in (45) looks like (with possible multiplicity $n$)
\beq
((\chi -z_i)(\bar\chi - \bar z_i))^n =:\zeta^n \bar\zeta^n
\eeq
which implies for the above expression (44) 
\beq
\frac{i}{2}(M_{,\chi}\chi_{,l}
- M_{,\bar\chi}\bar\chi_{,l}) \sim \frac{in}{2}\frac{\bar \zeta
\chi_{,l} - \zeta \bar\chi_{,l}}{\zeta \bar\zeta} + \, \ldots
\eeq
where the remainder is regular at $\zeta =0$. The above singularity may be
compensated by the pure gauge factor
\beq
\Lambda = -n \, {\rm atan} \frac{i(\zeta - \bar\zeta)}{\zeta + \bar\zeta} .
\eeq
Indeed ($\zeta_{,l} \equiv \chi_{,l}$),
\beq
\Lambda_{,l} = -\frac{in}{2}\frac{\bar \zeta
\zeta_{,l} - \zeta \bar\zeta_{,l}}{\zeta \bar\zeta}
\eeq
precisely cancels the singular term (47). The spinor in (32) 
is multiplied by the
gauge factor $\exp (i\Lambda)$. This factor is single-valued only if the
order $n$ of the zero of $PQ' - P'Q$ is integer, because $\Lambda$ 
in (48) 
is a multiply-valued function. This implies that both $P$ and $Q$ are
polynomials and, therefore,  restricts all zeros and poles of $R(z)$ 
to integer orders. If we demand, in addition, that the Hopf index (and,
consequently, the Chern--Simons action) is finite, then $R(z)$ is restricted
to the rational maps, as stated above. 

In fact, this is not yet the whole story about singularities in 
$\bar A^{(M)}_l$. The point is that the expression
\beq
\frac{i}{2}\frac{\bar \chi \chi_{,l} - \chi \bar\chi_{,l}}{\chi\bar\chi}
\eeq
is singular in the limit $\chi \to \infty$ as well, as may be checked 
easily. Further, the derivatives of the gauge factors, (48), for all the
zeros  of (45) produce this expression (50) for $\chi\to\infty$,
because $\lim_{\chi \to \infty}\zeta =\chi$. As there are $p+q-1$ zeros
(including multiplicities),  the sum of all $\Lambda_{,l}$ in (49) behaves
in the limit $\chi\to\infty$ as
\beq
\lim_{\chi\to\infty}\sum_{\rm zeros, poles} \Lambda_{,l}=
-(p+q-1)\frac{i}{2}\frac{\bar \chi \chi_{,l} - 
\chi \bar\chi_{,l}}{\chi\bar\chi} .
\eeq
In addition, eq. (44) produces a term that behaves in the limit $|\chi| \to
\infty$ as
\beq
-(p-q-1)\frac{i}{2}\frac{\bar \chi \chi_{,l} - 
\chi \bar\chi_{,l}}{\chi\bar\chi}
\eeq
as may be checked easily. Therefore, alltogether we have to compensate
(in the limit $|\chi |\to\infty$)
\beq
-2(p-1)\frac{i}{2}\frac{\bar \chi \chi_{,l} - 
\chi \bar\chi_{,l}}{\chi\bar\chi}
\eeq
by
an additional gauge transformation, without introducing further singularities
at $\chi =0$. 

Fortunately this is possible for the following reason. If we were to 
compensate (53) by the full gauge function
\beq
\Lambda = 2(p-1){\rm atan} \frac{i(\chi - \bar\chi)}{\chi + \bar\chi}
=-2(p-1)\sigma
\eeq
(where $\sigma$ is the phase of $\chi$ given in (16)), this would introduce 
a singularity at $\chi =0$. However, $\sigma$ is the sum of two terms
$\sigma = \sigma^{(1)} + \sigma^{(2)}$,
\beq
\sigma^{(1)}= {\rm atan} \frac{x_2}{x_1} \, , \quad 
\sigma^{(2)}= {\rm atan} \frac{1-r^2}{2x_3}
\eeq
where $\sigma^{(1)}_{,l}$ is singular at $\chi =0$ and $\sigma^{(2)}_{,l}$
is singular at $\chi = \infty$. Therefore, we may cancel the singularity
of (53) at $|\chi |=\infty$ without introducing further singularities by
performing an additional gauge transformation using only $\sigma^{(2)}$,
\beq
\Lambda =- 2(p-1)\sigma^{(2)} .
\eeq  

\section{Summary}

We have shown that, in the presence of the fixed prescribed background
magnetic field (22), there is an infinite number of fully 
three-dimensional solutions to the system of equations (23) and (24). These 
solutions are given by the set of Hopf maps (29), where the standard Hopf map 
(15) is followed by an arbitrary rational map (3); i.e., these solutions are
Hopf instantons.

Before closing, we want to briefly discuss some aspects of our solutions.
Firstly, one might ask whether the background gauge field (22) (and the 
corresponding magnetic field), which is crucial for our solutions to exist,
admits some further interpretation. Indeed, as was already explained in
\cite{hoin}, this background field may either be related to spin 1/2
solutions of the Dirac equation (23), or it may be interpreted as a spin 
connection term in the Dirac operator on a conformally flat three-manifold
with torsion, rather than as a background gauge field. For details we refer to
\cite{hoin}.

Secondly, we want to mention that it is possible to count the number of
solutions (43) for a given Hopf index $N = w^2$. 
As a solution is characterised by a
rational map (3), which is the ratio of two polynomials, the complex
coefficients
$P_i$ and $Q_i$ of the two polynomials $P(z)=\sum P_i z^i$ and $Q(z)=
\sum Q_i z^i$ parametrise these solutions.  
However, not each distinct rational map $R(z)$ leads to a different solution
$M$, (43). In fact, $M$ is invariant under the $SU(2)$ transformation
\beq
R(z) \ra \frac{aR(z) +b}{-\bar b R(z) +\bar a} \, , \quad
|a|^2 + |b|^2 =1 \, , \quad a,b\in {\rm\bf C} .
\eeq
Geometrically, this transformation corresponds to an $SO(3)$ rotation of
the target $S^2$. Therefore, we should parametrise independent solutions
$M$ by the set of punctured rational maps that leave one point invariant.
If we choose the invariant point to be the south pole, i.e., $R(z=\infty)
=\infty$, then this implies for the degrees $p$ and $q$ of the polynomials
$P$ and $Q$ and for the Hopf index $N$
\beq
N=p^2 > q^2 .
\eeq
With this restriction, a general rational map $R(z)$ with fixed $p$ has 
$4p$ real parameters, therefore we find a solution space of dimension 
$4p$ for Hopf index $N=p^2$ for our class of solutions (43).

It should be mentioned at this point that this classification of the space
of solutions is completely analogous to the case of the self-dual
Jackiw--Pi model. In this model a non-relativistic scalar field is
included in an Abelian Chern--Simons theory, and two-dimensional solitonic
solutions are found to exist \cite{JaPi1,JaPi2,Dun1,Horv1}. These soliton
solutions obey the Liouville equation (8), however, in coordinate space
rather than in target space. They are classified in a way that is completely 
analogous to our discussion above \cite{Horv1}, 
with the magnetic flux as the topological
quantity (where the magnetic flux is equal to $4\pi$ times
the degree $p$ of the rational map).

In fact, there is another relation to the self-dual Jackiw--Pi model. 
When our equations of motion (23) and (24) are dimensionally reduced by
assuming independence of $x_3$ and by setting $A_3 \equiv 0$, then the
resulting equations of motion are precisely the equations of motion of
the self-dual Jackiw--Pi model that we just described. These matters will be 
discussed in more detail elsewhere.

Finally we want to point out that our findings in no way imply that we
have already exhausted the space of solutions. All our solutions are
based on the simplest Hopf map $\chi$, (15), where the pre-image of a point
$\chi = {\rm const}$ is the simplest possible knot (the unknot). It is
perfectly possible that by starting from more complicated Hopf maps,
e.g., with more complicated knots as pre-images, one may find more
solutions. This question is subject to further investigation.    

\section{Acknowledgments}
The authors thank M. Fry for helpful discussions. In addition,
CA thanks R. Jackiw for useful conversations, and especially for pointing
out the relation to the Jackiw--Pi model.
CA was supported by a Forbairt Basic Research Grant during part of the
work.
BM gratefully acknowledges financial support from the Training and 
Mobility of Researchers scheme (TMR no. ERBFMBICT983476).


\begin{thebibliography}{999999}
\bibitem{hoin}
C. Adam, B. Muratori, C. Nash, hep-th/9909189
\bibitem{Ran1}
A. F. Ranada, J. Phys. A25 (1992) 1621
\bibitem{FN1}
L. Faddeev and A. Niemi, Nature 387 (1997) 58
\bibitem{FN2}
L. Faddeev and A. Niemi, hep-th/9705176
\bibitem{GH1}
J. Gladikowski and M. Hellmund, Phys. Rev. D56 (1997) 5194
\bibitem{BS1}
R. Battye and P. Sutcliffe, hep-th/9811077
\bibitem{BS2}
R. Battye and P. Sutcliffe, Phys. Rev. Lett. 81 (1998) 4798
\bibitem{AFZ}
H. Aratyn, L. A. Ferreira and A. H. Zimerman, Phys. Rev. Lett. 83 (1999)
1723
\bibitem{DJT}
S. Deser, R. Jackiw and S. Templeton, Ann. Phys. 140 (1982) 372
\bibitem{JLW}
R. Jackiw, K. Lee and E. J. Weinberg, Phys. Rev. D42 (1990) 3488
\bibitem{Wan1} 
R. Wang, Comm. Math. Phys. 137 (1991) 587
\bibitem{JaPi1}
R. Jackiw and S.-Y. Pi, Prog. Theor. Phys. (Suppl.) 107 (1992) 1
\bibitem{JaPi2}
R. Jackiw and S.-Y. Pi, Phys. Rev. Lett. 64 (1990) 2969; 66 (1991) 2682;
Phys. Rev. D42 (1990) 3500
\bibitem{Dun1}
G. Dunne, hep-th/9902115
\bibitem{ZHK}
S. C. Zhang, T. H. Hansson and S. Kivelson, Phys. Rev. Lett. 62 (1989) 82
\bibitem{Jain}
J. K. Jain, Phys. Rev. Lett. 63 (1989) 199
\bibitem{JaPi3}
R. Jackiw and S.-Y. Pi, hep-th/9911072 
\bibitem{LoYa}
M. Loss and H.-Z. Yau, Comm. Math. Phys. 104 (1986) 283
\bibitem{Zero}
C. Adam, B. Muratori and C. Nash, Phys. Rev. D60 (1999) 125001 
\bibitem{HMS}
C. Houghton, N. Manton and P. Sutcliffe, Nucl. Phys. B510 (1998) 507
\bibitem{Horv1}
P. A. Horvathy, hep-th/9903116


\end{thebibliography}
\end{document}